\documentclass{elsart5p}

\usepackage{graphics}
\usepackage{graphicx}
\usepackage{amssymb}

\begin{document}

\begin{frontmatter}



\title{Thermal expansion and pressure effect in $MnWO_4$}
%

\author[AA]{R. P. Chaudhury\corauthref{Name1}},
\ead{rajit.chaudhury@uh.edu}
\author[AA]{F. Yen},
\author[AA]{C. R. dela Cruz},
\author[AA]{B. Lorenz},
\author[AA]{Y. Q. Wang},
\author[AA]{Y. Y. Sun},
\author[AA,BB,CC]{C. W. Chu}

\address[AA]{TCSUH and Department of Physics, University of Houston, Houston, Texas, 77204-5002, USA}
\address[BB]{Lawrence Berkeley National Laboratory, 1 Cyclotron Road, Berkeley, California, 94720, USA }
\address[CC]{Hong Kong University of Science and Technology, Hong Kong, China}

\corauth[Name1]{Corresponding author. Tel: (713) 743-8314 fax: (713)
743-8201}

\begin{abstract}
$MnWO_4$ has attracted attention because of its ferroelectric
property induced by frustrated helical spin order. Strong
spin-lattice interaction is necessary to explain ferroelectricity
associated with this type of magnetic order. We have conducted
thermal expansion measurements along the $a$, $b$, $c$ axes
revealing the existence of strong anisotropic lattice anomalies at
$T_1$=7.8 K, the temperature of the magnetic lock-in transition into
a commensurate low-temperature (reentrant paraelectric) phase. The
effect of hydrostatic pressure up to 1.8 GPa on the FE phase is
investigated by measuring the dielectric constant and the FE
polarization. The low-temperature commensurate and paraelectric
phase is stabilized and the stability range of the ferroelectric
phase is diminished under pressure.
\end{abstract}

\begin{keyword}
$MnWO_4$; thermal expansion; pressure effect; multiferroics
\PACS 75.30.-m,75.30.Kz,75.50.Ee,77.80.-e,77.84.Bw
\end{keyword}

\end{frontmatter}

Multiferroic magnetoelectric compounds exhibit the coexistence of
ferroelectric (FE) and magnetic orders in some temperature range.
The mutual correlation between these orders is of fundamental
physical interest and it bears the potential for future applications
utilizing the magnetoelectric effect in which the magnetization (FE
polarization) is controlled by internal or external electric
(magnetic) fields \cite{kimura:03,goto:04}. Recently, this property
has been observed in $MnWO_4$ in a phase with an incommensurate (IC)
helical spin density wave \cite{taniguchi:06}. $MnWO_4$ crystalizes
in the wolframite structure (monoclinic space group P2/c). Below 15
K competing magnetic exchange interactions result in a high level of
magnetic frustration with several magnetically ordered states
quasi-degenerated in energy. As a consequence, $MnWO_4$ undergoes
three successive magnetic transitions, antiferromagnetic (AFM) order
of the Mn-spins with an IC sinusoidal spin modulation appears at
$T_N$=13.5 K (AF3 phase) followed by an elliptical IC magnetic order
below $T_2$=12.6 K (AF2 phase) and a commensurate collinear magnetic
phase below $T_1$=7.8K (AF1 phase) \cite{lautenschlager:93}.
Ferroelectricity was observed in the AF2 phase only and it can
qualitatively be explained by the loss of inversion symmetry due to
the helical magnetic order and a strong spin-lattice coupling
\cite{mostovoy:06}.

\begin{figure}
\begin{center}
\includegraphics[angle=-90,width=0.45\textwidth]{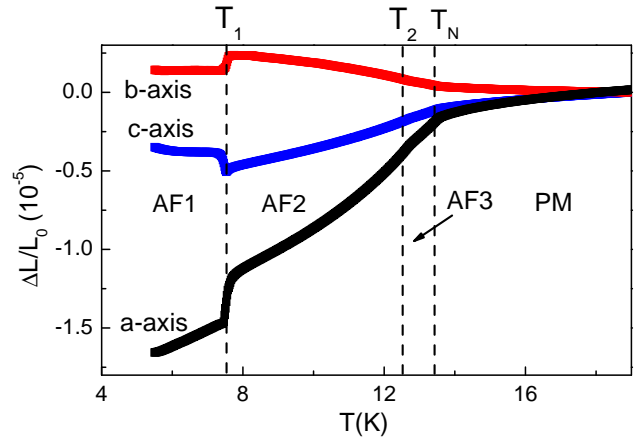}
\end{center}
\caption{Thermal expansion of the lattice parameters of $MnWO_4$.}
\label{fig1}
\end{figure}

The magnetic phase transitions are also visible in anomalies of the
specific heat, the dielectric constant, and the magnetic
susceptibility \cite{arkenbout:06}.

The coupling between AFM and FE orders observed in $MnWO_4$ must be
mediated by strong spin-lattice interactions. The existence of such
spin-lattice coupling can be experimentally proven by detecting the
strain of the lattice by high-resolution thermal expansion
measurements \cite{delacruz:06, chaudhury:07}. The macroscopic
lattice strain along the principal crystallographic orientations,
$a$, $b$, and $c$, is measured employing a high-resolution
capacitance dilatometer. The results shown in Fig.1 reveal clear
anomalies of all three lattice constants at $T_N$ (change of slope
of all lattice parameters) and at $T_1$, the transition into the
re-entrant paraelectric state (sharp step-like changes of $a$, $b$,
$c$).

\begin{figure}
\begin{center}
\includegraphics[angle=-90,width=0.44\textwidth]{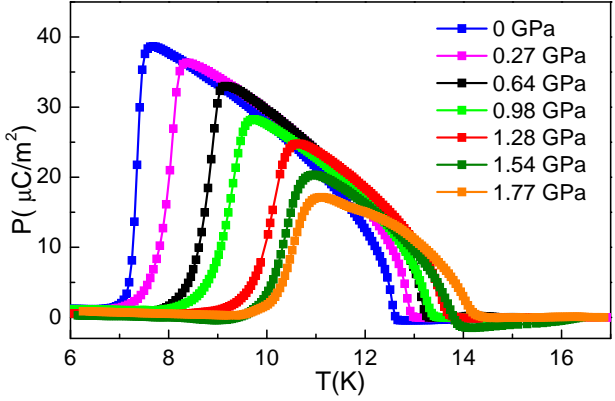}
\end{center}
\caption{FE polarization of $MnWO_4$ at various pressures.}
\label{fig2}
\end{figure}

Is is important to notice the strong anisotropy of the lattice
strain at $T_1$. The relative changes of $a$, $b$, $c$ and the
volume $V$ at $T_1$ are: $\Delta a/a=-3\times10^{-5}$, $\Delta
b/b=-1\times10^{-5}$, $\Delta c/c=1\times10^{-5}$ and $\Delta
V/V=-3\times10^{-5}$, where $\Delta V=V(T<T_1)-V(T>T_1)$. The volume
of the AF1 phase is smaller than that of the AF2 phase. The
discontinuous volume change across $T_1$ proves the first order
nature of this phase transition. The anisotropic strain observed in
the thermal expansion measurements is associated with the magnetic
anisotropy and the peculiar changes of the magnetic order parameter
at the various phase transitions. The details of the magnetic
structure was revealed in neutron scattering experiments
\cite{lautenschlager:93}. In the collinear AF1 and AF3 phases the
spins are aligned with the easy axis of magnetization that lies in
the $a$-$c$ plane at an angle of 37$^\circ$ with the $a$-axis. In
the helical AF2 phase the spin has a component along $b$. The
propagation vector $\overrightarrow{q}=(-0.214, 0.5, 0.457)$ of the
AF2 and AF3 phases abruptly changes at $T_1$ to the commensurate
$\overrightarrow{q}=(-0.25, 0.5, 0.5)$ in the AF1 phase. The
re-alignment of the spins with the easy axis and the sudden locking
of the magnetic modulation with the lattice causes the large
decrease of the $a$-axis and the associated volume change. The
magnetization along the easy axis also shows the largest change at
$T_1$ \cite{arkenbout:06}.

The strong magnetoelastic effects demonstrated above (Fig. 1) imply
a large sensitivity of the AFM and FE orders with respect to lattice
strain affecting the interatomic distances and the magnetic exchange
interaction parameters. Therefore, external pressure, $p$, can be
used to modify the intrinsic magnetic interactions. This is
complimentary to the application of external magnetic fields that
directly couple to the spins. Of particular interest is the
stability of the ferroelectric phase (AF2) with a spontaneous
polarization that can be measured by the pyroelectric current
method. Fig. 2 shows the FE polarization at different hydrostatic
pressures up to 1.8 GPa. Both $T_1$ and $T_2$ increase with $p$ but
$T_1$ increases at a faster rate diminishing the stability range of
the FE phase at higher $p$. The FE polarization is suppressed by
pressure, in particular at the low-temperature end of the FE phase.
The resulting phase diagram is shown in Fig. 3. The critical
temperature of the AF3 phase, $T_N$, could not be resolved under
pressure. Extrapolating the $p$-dependence of $T_1$ and $T_2$ we
determine the critical pressure above which the FE state becomes
unstable as $p_c \approx$ 4 GPa. The strong increase of $T_1$ can be
explained by the volume effect. Since the commensurate AF1 phase has
a sizably smaller volume than the FE AF2 phase pressure will
stabilize the AF1 phase and increase $T_1$. The current phase
diagram can be compared to the $p$-$T$ phase diagram of $Ni_3V_2O_8$
that shows a similar sequence of magnetic and FE transitions below
10 K \cite{chaudhury:07}. In the latter case the critical pressure
suppressing ferroelectricity was found to be much lower, 1.64 GPa.
$Ni_3V_2O_8$ is very different in terms of lattice structure and
symmetry. Unlike $MnWO_4$, the lattice of $Ni_3V_2O_8$ is highly
anisotropic and the magnetic frustration is not only due to
competing exchange interactions but also due to the geometry of the
Kagome staircase structure of the magnetic Ni-sublattice. This could
be the origin for the substantially higher sensitivity of the
helical spin structure to the applied pressure.

\begin{figure}
\begin{center}
\includegraphics[angle=-90,width=0.44\textwidth]{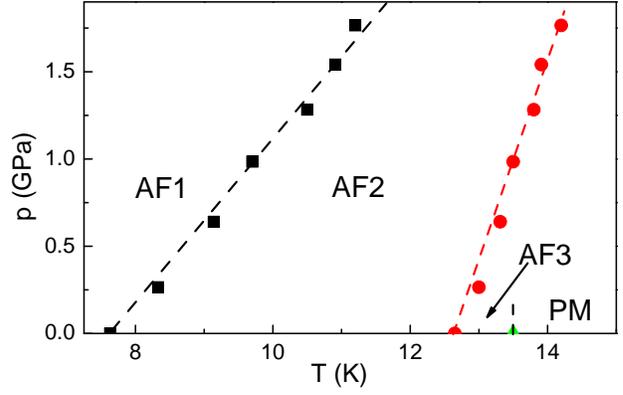}
\end{center}
\caption{Pressure-temperature phase diagram of $MnWO_4$.}
\label{fig3}
\end{figure}

This work is supported in part by the T.L.L. Temple Foundation, the
J. J. and R. Moores Endowment, and the State of Texas through TCSUH
and at LBNL through the US DOE.

\bibliographystyle{phpf}


\end{document}